\documentclass[aps, pre, preprint, groupedaddress, amsmath, amssymb]{revtex4-2}

\usepackage{graphicx}
\usepackage{bm}
\usepackage{siunitx}
\usepackage{hyperref}

\begin{document}

\title{GPU-Accelerated Sequential Monte Carlo for Bayesian Spectral Analysis}

\author{Tomohiro Nabika}
\affiliation{a.s.ist Inc.}
\author{Yui Hayashi}
\affiliation{a.s.ist Inc.}
\author{Masato Okada}
\affiliation{Graduate School of Frontier Sciences, The University of Tokyo}

\date{March 23, 2026}

\begin{abstract}
Bayesian spectral deconvolution provides a data-driven framework for mathematical model selection and parameter estimation from spectral data.
Although highly versatile, it becomes computationally expensive as the number of model parameters, data points, and candidate models increases, often rendering practical applications infeasible.
We propose a GPU-accelerated approach in which a sequential Monte Carlo sampler (SMCS) is run in parallel on a GPU to perform Bayesian model selection of the number of spectral peaks and Bayesian estimation of peak-function parameters.
Numerical experiments demonstrate that the GPU-parallelized SMCS achieves speedups exceeding $500\times$ over CPU-parallelized replica exchange Monte Carlo (REMC).
The method is validated on artificial data designed to emulate X-ray photoelectron spectroscopy (XPS) and X-ray diffraction (XRD) measurements, as well as on real experimental spectra.
As measurement techniques such as microscopic spectroscopy and in-situ methods continue to drive rapid growth in the volume of spectral data, the proposed approach offers a practical computational foundation for advanced analysis of individual datasets.
\end{abstract}

\maketitle

\section{\label{sec:introduction}Introduction}

The development and advancement of functional materials underpin a wide range of modern industrial sectors~\cite{de2019new, panwar2020material}.
Designing and evaluating these materials requires the analysis of measurement data that encode structural and chemical-state information.
X-ray diffraction (XRD) and X-ray photoelectron spectroscopy (XPS) are among the most widely used techniques, yielding diffraction and spectral data that are routinely employed to characterize crystal structures, identify phases, and assess surface compositions and chemical bonding states~\cite{davel2025machine, krishna2022review}.

A prevalent approach to spectral data analysis constructs a forward model in which peaks are represented by basis functions---Gaussian, Lorentzian, or Voigt profiles---and fits model parameters to the data by minimizing a cost function such as the least-squares residual~\cite{wojdyr2010fityk, toby2013gsas, brehm2023pyramangui}.
The resulting peak count and parameter values provide direct insight into material properties, crystal structures, chemical compositions, and bonding states.

Conventionally, the number of basis functions is chosen by the analyst based on spectral derivatives or physical knowledge of the sample.
Gradient-based optimizers---steepest descent, quasi-Newton methods---are then applied for parameter fitting~\cite{wojdyr2010fityk, toby2013gsas, brehm2023pyramangui}.
Because cost functions in spectral analysis typically possess multiple local optima, these algorithms frequently converge to solutions that depend sensitively on the initial conditions, necessitating repeated trial-and-error adjustment of starting values until a plausible result is obtained.
Global search methods such as genetic algorithms and simulated annealing have been employed to mitigate this issue, yet they yield only point estimates and offer no systematic means of quantifying estimate reliability.
Conventional spectral analysis thus suffers from three intertwined problems: reliance on subjective judgment, susceptibility to local optima, and the absence of rigorous uncertainty quantification.

Bayesian inference offers a principled resolution of these difficulties.
Several data-driven methods based on Bayesian spectral deconvolution have been proposed~\cite{nagata2012bayesian, nagata2019bayesian, tokuda2017simultaneous, murakami2024bayesian, machida2021bayesian}.
By computing the posterior probability over forward models and their parameters, these methods enable the number of basis functions and the model parameters to be estimated in a purely data-driven fashion.
Employing replica exchange Monte Carlo (REMC, also known as parallel tempering)~\cite{hukushima1996exchange, earl2005parallel} for posterior sampling ensures stable convergence to the global optimum, while credible intervals and standard deviations derived from the posterior provide a rigorous measure of estimate reliability.

The principal obstacle to broader adoption of Bayesian spectral analysis is the large computational cost of the sampling algorithms.
One approach to acceleration reformulates the problem as maximum \textit{a posteriori} (MAP) point estimation~\cite{matsumura2024maximum}.
Although fast and capable of incorporating explicit prior knowledge, this approach---like other existing methods---cannot compute marginal likelihoods for Bayesian model selection and does not furnish quantitative uncertainty estimates.
Alternative strategies that sample the peak count jointly with the model parameters have also been proposed to streamline Bayesian model selection~\cite{kawashima2019bayesian, okajima2021fast}.
In the method of Kawashima \textit{et al.}~\cite{kawashima2019bayesian}, however, no theoretical efficiency bound is provided, and the observed speedup is only about twofold.
Okajima \textit{et al.}~\cite{okajima2021fast} estimate a theoretical upper bound of $O(K_\text{max}/2)$ for the efficiency gain, where $K_\text{max}$ is the maximum candidate peak count, yet the speedup observed in numerical experiments was limited to roughly sixfold.

In this work, we propose accelerating posterior sampling in spectral data analysis by deploying a sequential Monte Carlo sampler (SMCS)~\cite{doucet2001introduction, del2006sequential, betz2016transitional, chopin2020introduction} on a GPU.
The SMCS escapes local optima through resampling, achieving---like REMC---stable convergence to the global optimum.
A critical distinction lies in the available parallelism.
REMC parallelizes over inverse-temperature chains, whose number is typically $O(10\text{--}10^2)$, whereas the SMCS parallelizes over particles, parameters, and data points---dimensions that commonly reach $O(10^4\text{--}10^6)$~\cite{hendeby2010particle}.
Moreover, because SMC sampling involves frequent tensor operations, GPU-parallel implementations of SMC-based Bayesian inference have already proven effective in diverse scientific domains~\cite{lee2010utility, yallup2025particle, white2023smcdeblending, speich2021sequential, lubbe2022bayesian}.
We bring this GPU-parallel strategy to spectral data analysis, thereby alleviating the computational bottleneck that has limited existing Bayesian approaches.

We benchmark the GPU-parallelized SMCS against CPU-parallelized REMC using both artificial spectral data and real XRD and XPS measurements, with convergence of the Bayesian free energy serving as the primary comparison metric.

The remainder of this paper is organized as follows.
Section~\ref{sec:bayesian} formulates Bayesian spectral analysis and describes the sampling algorithms.
Section~\ref{sec:experiments} presents numerical experiments on both artificial and real data.
Section~\ref{sec:discussion} discusses the results, and Sec.~\ref{sec:conclusion} gives concluding remarks and an outlook for future work.

\section{\label{sec:bayesian}Bayesian spectral analysis}

\subsection{\label{sec:formulation}Formulation}

Spectral deconvolution constructs a mathematical model from basis functions such as Gaussian, Lorentzian, or Voigt profiles and estimates the model parameters by minimizing a cost function evaluated against measured data.
Let $D=\{x_i, y_i\}_{i=1}^N$ denote the measured data, $\theta$ the model parameters, $g_k(x;\theta_k)$ the $k$th basis function, and $b(x)$ a baseline.
The forward model $f(x;\theta)$ is then
\begin{equation}
  f(x;\theta) = \sum_{k=1}^K g_k(x; \theta_k) + b(x). \label{eq:forward_model}
\end{equation}

In Bayesian spectral deconvolution, the peak count $K$ and the model parameters $\theta$ are estimated by computing their posterior probabilities from the measured data.
The posterior probability of $K$ is
\begin{align}
  p(K| D) &= \frac{p(D | K)\,p(K)}{p(D)} \nonumber\\
  &= \frac{\exp\bigl(-F(K)\bigr)\,p(K)}{\sum_{k}p(D|k)\,p(k)},
\end{align}
where $F(K)$ is the Bayesian free energy, defined by
\begin{equation}
  F(K) = -\log \int p(D | K, \theta)\, p(\theta|K)\, d\theta. \label{eq:free_energy_def}
\end{equation}

For a given $K$, the posterior distribution of the model parameters $\theta$, written $p(\theta|D) \equiv p(\theta|D,K)$, takes the form
\begin{align}
  p(\theta|D) &= \frac{p(D \mid \theta)\,p(\theta)}{\int p(D\mid\theta)\,p(\theta)\,d\theta}
  \label{eq:posterior_theta}\\
  &= \frac{p(D \mid \theta)\,p(\theta)}{Z(D)},
  \label{eq:posterior_theta_def}
\end{align}
where $Z(D)$ is the normalizing constant, commonly referred to as the marginal likelihood.

\subsection{\label{sec:sampling}Sampling algorithms}

Bayesian spectral deconvolution requires samples from the posterior distribution in Eq.~\eqref{eq:posterior_theta}.
Because the parameter vector $\theta$ is generally high-dimensional and $p(\theta|D)$ typically possesses multiple local modes, the sampling algorithm must be capable of escaping local optima and reliably reaching the global optimum.
Below, we describe REMC (parallel tempering), which has been widely used for this purpose, and the SMCS adopted in the present work.

\subsubsection{\label{sec:remc}Replica exchange Monte Carlo}

REMC facilitates transitions between local modes of a multimodal posterior by introducing a sequence of tempered distributions in which the likelihood is softened by an inverse temperature $\beta$~\cite{hukushima1996exchange, earl2005parallel}.
For a ladder $\beta_0=0<\beta_1<\cdots<\beta_L=1$, each replica $\ell$ performs MCMC updates targeting
\begin{equation}
  p_{\beta_\ell}(\theta\mid D)\ \propto\ p(D\mid \theta)^{\beta_\ell}\,p(\theta).
\end{equation}
At regular intervals, a swap of states $\theta_\ell \leftrightarrow \theta_{\ell+1}$ between adjacent replicas $(\ell,\ell{+}1)$ is proposed and accepted with probability
\begin{equation}
  \alpha
  = \min\!\left\{
    1,\
    \exp\!\Bigl[(\beta_{\ell+1}-\beta_{\ell})
    \left(E_{\ell+1}-E_\ell\right)\Bigr]\right\},
\end{equation}
where $E_\ell=-\log p(D\mid\theta_\ell)$.
The Bayesian free energy [Eq.~\eqref{eq:free_energy_def}] can be computed from the REMC samples as~\cite{nagata2012bayesian}
\begin{align}
  Z(\beta{=}1) &= \prod_{\ell=0}^{L-1} \frac{Z(K, \beta_{\ell+1})}{Z(K, \beta_{\ell})} \nonumber\\
  &= \prod_{\ell=0}^{L-1} \left\langle \exp\!\left[-(\beta_{\ell+1} - \beta_{\ell})\, E_\ell \right] \right\rangle_{p_{\beta_\ell}(\theta_\ell|D, K)}.
  \label{eq:final_marginal_likelihood}
\end{align}
Broad exploration at high temperatures propagates through the exchange mechanism to the target posterior at $\beta{=}1$, thereby improving mixing relative to single-temperature MCMC.
The natural unit of parallelism in REMC is the number of inverse temperatures $L$, which typically ranges from tens to a few hundred, making the algorithm well suited for CPU-based parallelization.

\subsubsection{\label{sec:smc}Sequential Monte Carlo sampler}

The sequential Monte Carlo (SMC) sampler approximates a sequence of distributions $\{\pi_\ell(\theta)\}_{\ell=0}^L$ leading to the target by maintaining a large set of weighted particles and iterating weight updates, resampling, and transition moves until the particles converge to $\pi_L$~\cite{doucet2001introduction, del2006sequential, chopin2020introduction}.
We construct the tempered sequence from the prior to the posterior using inverse temperatures $\beta_0=0<\beta_1<\cdots<\beta_L=1$:
\begin{equation}
  \pi_\ell(\theta)\ \propto\ p(D\mid\theta)^{\beta_\ell}\,p(\theta).
  \label{eq:smc_tempered_sequence}
\end{equation}
Given particles $\{\theta_{\ell-1}^{(i)}\}_{i=1}^T$ from the previous step, the incremental importance weights are
\begin{equation}
  w_\ell^{(i)}\ \propto\ p(D\mid \theta_{\ell-1}^{(i)})^{\beta_\ell-\beta_{\ell-1}}.
  \label{eq:smc_incremental_weight}
\end{equation}
Particles are resampled according to the normalized weights $\tilde{w}_\ell^{(i)}=w_\ell^{(i)}/\sum_{j=1}^T w_\ell^{(j)}$, and each particle is then moved by an MCMC kernel that leaves $\pi_\ell$ invariant, restoring particle diversity.
Denoting the normalizing constant at inverse temperature $\beta_\ell$ by $Z_\ell$, the ratio of successive constants is estimated as the mean unnormalized weight:
\begin{equation}
  \frac{Z_\ell}{Z_{\ell-1}}\approx \frac{1}{T}\sum_{i=1}^T w_\ell^{(i)},\qquad
  \log Z_L = \sum_{\ell=1}^L \log\!\left(\frac{1}{T}\sum_{i=1}^T w_\ell^{(i)}\right),
  \label{eq:smc_marginal_likelihood}
\end{equation}
from which the Bayesian free energy $F=-\log Z_L$ is obtained~\cite{del2006sequential}.
For model selection, $F(K)$ is computed in this manner for each candidate peak count $K$ and the model with the smallest $F(K)$ is selected.
Because the SMCS typically operates with $T=O(10^4\text{--}10^6)$ particles, likelihood evaluations and weight updates can be parallelized across the particle, parameter, and data dimensions, naturally exploiting the massively parallel architecture of modern GPUs.

\subsubsection{\label{sec:waste_free}Waste-free SMC}

In the standard SMCS, particle diversity degrades unless a sufficient number of MCMC transitions are performed after each resampling step, yet increasing the number of transitions raises the computational cost.
Dau and Chopin~\cite{Dau2022_WasteFree} resolve this trade-off with waste-free SMC.

Rather than resampling all $T$ particles and applying $n$ MCMC transitions to each---totaling $Tn$ transitions---waste-free SMC resamples only $S = T/n$ particles (where $n$ is the number of MCMC steps) and applies $n$ transitions to each of these $S$ particles.
All intermediate MCMC states are retained, yielding a particle set of size $S \times n = T$ that is passed to the next step.
The total number of MCMC operations is thus approximately $T$, independent of $n$, allowing $n$ to be increased without additional cost.
In practice, however, $S$ must not be chosen too small, as this would amplify resampling variance.
We adopt waste-free SMC throughout this work.

\subsubsection{\label{sec:mcmc_kernel}MCMC kernel}

Both REMC and the SMCS employ a component-wise random-walk Metropolis--Hastings algorithm~\cite{metropolis1953equation, hastings1970monte} as the MCMC kernel.
Rather than updating all $d$ components of $\boldsymbol{\theta} = (\theta_1, \ldots, \theta_d)$ simultaneously, each component $\theta_i$ is updated individually in sequence.
Because each update involves only a one-dimensional proposal, the computational cost per step is low and the procedure maps naturally onto GPU hardware for large-scale parallelization.
An additional advantage is that the acceptance rate can be monitored independently for each component, facilitating per-component adaptive step-size tuning.

Step sizes are adjusted following Nabika \textit{et al.}~\cite{nabika2025_SEMC}.
In the SMCS, the step size at each temperature level is predicted explicitly from the histories of inverse temperatures $\beta_\ell$ and per-component acceptance rates.
In REMC, step sizes are adjusted on the fly via a Robbins--Monro algorithm~\cite{atchade2005adaptive}.

\section{\label{sec:experiments}Numerical experiments}

This section presents numerical experiments comparing CPU-parallelized REMC [hereafter REMC(CPU)] and GPU-parallelized SMCS [hereafter SMCS(GPU)].
The sample size of each method is progressively increased, and the convergence of the Bayesian free energy and the posterior distribution is tracked as a function of wall-clock time.
By varying the number of data points $N$, we also examine the scaling behavior of both methods to assess how effectively GPU parallelization copes with growing data size.
Details of the computational environment are given in Appendix~\ref{sec:computational_env}.

\subsection{\label{sec:artificial}Artificial data}

Before applying the method to real measurements, we validate it on artificially generated spectral data.
Artificial data are valuable for two reasons.
First, because the true model parameters are known, the estimation accuracy of the Bayesian free energy and the fidelity of the posterior distribution can be assessed quantitatively.
Second, the data size $N$ and model complexity can be varied freely, permitting a systematic study of computational scaling.

Throughout this subsection, the reference (baseline) value is taken from the SMCS(GPU) run with the largest particle count.
Free-energy convergence is quantified by the absolute error $|\Delta F|$ between each estimate and the baseline.
Posterior convergence is assessed through the errors in the posterior mean and the endpoints of the 95\% and 99\% credible intervals relative to the baseline.

\subsubsection{\label{sec:xrd_artificial}XRD data}
\paragraph{Experimental setup.}

The XRD artificial data analysis employs a reference-spectrum model for three TiO$_2$ polymorphs---rutile, anatase, and brookite.
Based on the forward model described in Appendix~\ref{sec:appendix_xrd} [Eq.~\eqref{eq:xrd_forward_model}], each phase $k$ is characterized by nine parameters: a $2\theta$ shift, asymmetry parameter $\alpha_k$, Gaussian--Lorentzian mixing ratio $r_k$, Gaussian width parameters $(u_k, v_k, w_k)$, Lorentzian width parameters $(s_k, t_k)$, and peak intensity $A_k$.
Together with four background parameters (amplitude $a$, width $\sigma_{\mathrm{bg}}$, mixing ratio $r_{\mathrm{bg}}$, and constant offset $b$), the total number of model parameters is $3 \times 9 + 4 = 31$.

\begin{figure}
  \includegraphics[width=\columnwidth]{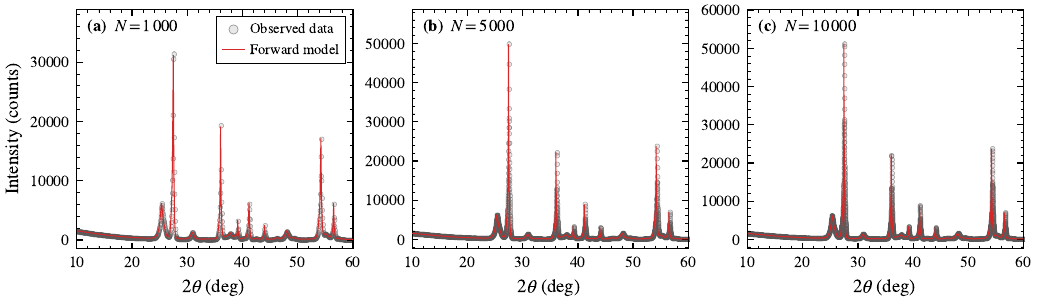}
  \caption{\label{fig:xrd_artificial_data}Artificial XRD data and the corresponding forward model for (a) $N=1\,000$, (b) $5\,000$, and (c) $10\,000$. Blue dots represent the observed data, and red solid lines represent the forward model evaluated at the true parameter values. Peaks from the reference spectra of the three TiO$_2$ phases (rutile, anatase, brookite) are visible.}
\end{figure}

To systematically investigate how data size affects computational speed, experiments were conducted at $N = 1000$, $5000$, and $10\,000$.
For each $N$, artificial data were created by adding noise [Eq.~\eqref{eq:xrd_poisson}] to spectra generated from the known parameter values.
Figure~\ref{fig:xrd_artificial_data} compares the artificial observations with the true forward model for each data size.
Larger $N$ yields denser sampling and makes reproduction of the forward model easier, but increases the cost of each likelihood evaluation.

For REMC(CPU), the number of inverse temperatures was $L=44$ and the burn-in fraction was 50\% of the total MCMC steps.
The total number of MCMC steps (including burn-in) was set to $\{100, 300, 1000, 3000\}$ by multiplying a base unit of 100 by factors of $\{1, 3, 10, 30\}$.
For SMCS(GPU), the number of MCMC steps per temperature level was $n = 10$, and the particle count was varied from 100 to $10^6$ by multiplying a base unit of 100 by $\{1, 3, 10, 30, 100, 300, 1000, 3000, 10\,000\}$.
Each condition was run independently 100 times, and the mean and standard deviation of the Bayesian free energy and computation time were recorded.

\paragraph{Free-energy comparison.}

\begin{figure}
  \includegraphics[width=\columnwidth]{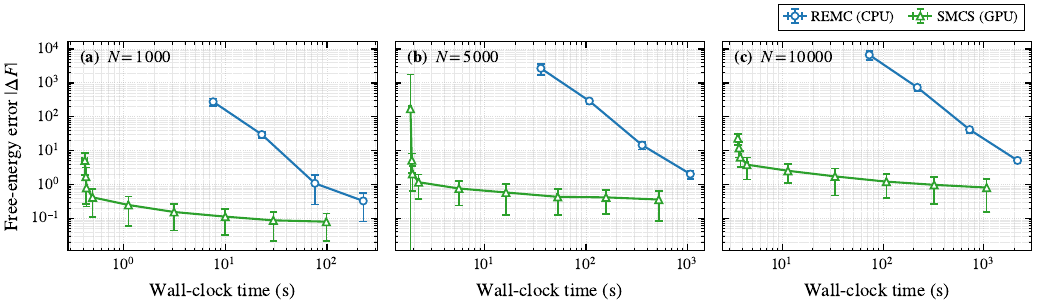}
  \caption{\label{fig:xrd_fe_error}Free-energy error $|\Delta F|$ versus computation time for XRD artificial data at (a) $N=1000$, (b) $5000$, and (c) $10\,000$. Blue circles: REMC(CPU); green triangles: SMCS(GPU).}
\end{figure}

Figure~\ref{fig:xrd_fe_error} shows the relationship between the free-energy error $|\Delta F|$ and computation time for each data size $N$.
The reference free energy is the trial-averaged value of SMCS(GPU) at the largest particle count.
Both axes use logarithmic scales; each data point represents the mean of 100 independent runs, with error bars indicating the standard deviation.

At every data size, SMCS(GPU) reaches a given free-energy accuracy in substantially less time than REMC(CPU).
Comparing the wall-clock time required to attain the same free-energy error as the largest REMC(CPU) run (3000 steps), the speedup factors are $316\times$ for $N{=}1000$, $547\times$ for $N{=}5000$, and $532\times$ for $N{=}10\,000$ (Table~\ref{tab:xrd_speedup}).

\begin{table}
  \caption{\label{tab:xrd_speedup}Speedup of SMCS(GPU) over REMC(CPU) for XRD artificial data at each data size $N$. The speedup is computed from the wall-clock time at which SMCS(GPU) attains the same $|\Delta F|$ as the largest REMC(CPU) run.}
  \begin{ruledtabular}
  \begin{tabular}{cccc}
    $N$ & REMC time (s) & SMCS time (s) & Speedup \\
    \hline
    1000  & 227.5  & 0.72 & $316\times$ \\
    5000  & 1066.2 & 1.95 & $547\times$ \\
    10000 & 2163.6 & 4.07 & $532\times$ \\
  \end{tabular}
  \end{ruledtabular}
\end{table}

The signed estimation error $F - F_{\mathrm{ref}}$ is plotted against computation time in Fig.~\ref{fig:xrd_fe_value} (Appendix~\ref{sec:appendix_fe_convergence}).
For both methods, the error converges to zero with increasing sample size, confirming that both samplers target the same distribution.

\paragraph{Scaling of the speedup with data size.}

As shown in Table~\ref{tab:xrd_speedup}, increasing $N$ from 1000 to 5000 raises the speedup from $316\times$ to $547\times$.
This occurs because the REMC(CPU) computation time grows nearly linearly with $N$, whereas SMCS(GPU) absorbs much of the increase through data-level parallelism, so that SMCS(GPU) benefits more from larger datasets.
Between $N{=}5000$ and $N{=}10\,000$, however, the speedup plateaus at approximately $530\times$.

\paragraph{Comparison of posterior credible intervals.}

\begin{figure}
  \includegraphics[width=\columnwidth]{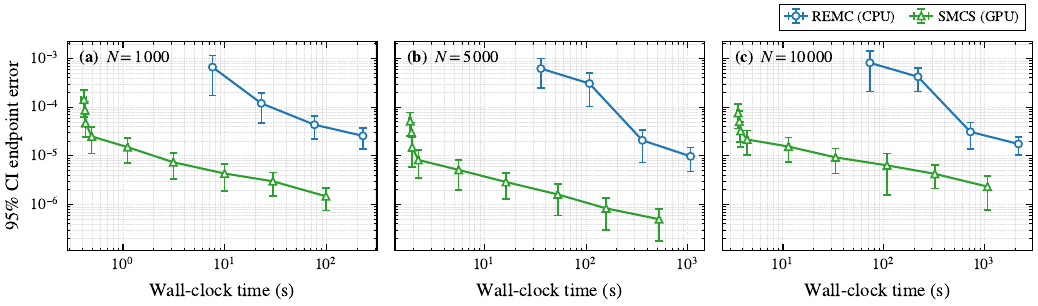}
  \caption{\label{fig:xrd_param_error}Sum of absolute endpoint errors in the 95\% credible interval for the rutile $2\theta$ shift parameter versus computation time (XRD artificial data). Results for (a) $N=1\,000$, (b) $5\,000$, and (c) $10\,000$ are shown. Blue circles: REMC(CPU); green triangles: SMCS(GPU).}
\end{figure}

Figure~\ref{fig:xrd_param_error} shows the convergence of the 95\% credible interval for the rutile-phase $2\theta$ shift parameter by plotting the summed absolute error of the upper and lower endpoints against computation time.
The baseline for each endpoint is the trial-averaged value at the largest SMCS(GPU) particle count ($10^6$ particles).
Consistent with the free-energy results, SMCS(GPU) reaches the same level of convergence in less wall-clock time than REMC(CPU).

\subsubsection{\label{sec:spectral_artificial}Spectral deconvolution model}
\paragraph{Experimental setup.}

The spectral deconvolution model employs a multi-peak model with Gaussian basis functions.
Each observation $y_i$ is generated as a superposition of $K$ Gaussian peaks with additive Gaussian noise:
\begin{equation}
  y_i = \sum_{k=1}^{K} A_k \exp\!\left(-\frac{b_k}{2}(x_i - \mu_k)^2\right) + \varepsilon_i, \quad \varepsilon_i \sim \mathcal{N}(0,\sigma^2).
\end{equation}
Each peak has three parameters---amplitude $A_k$, center $\mu_k$, and width $b_k$---giving $3K$ parameters in total.

To investigate how the number of peaks affects the speedup, experiments were performed at $K=3$, $10$, and $30$, with corresponding data sizes $N=300$ ($x\in[0,3]$), $1000$ ($x\in[0,10]$), and $3000$ ($x\in[0,30]$).
True parameter values and prior distributions are detailed in Appendix~\ref{sec:appendix_artificial_model}.
Figure~\ref{fig:spectral_artificial_data} shows the artificial observations together with the true forward model for each setting.

\begin{figure}
  \includegraphics[width=\columnwidth]{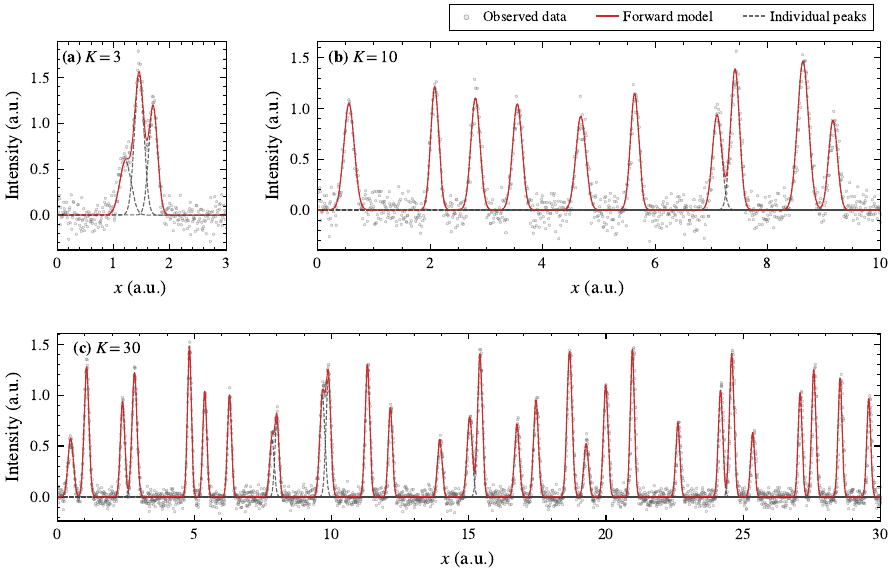}
  \caption{\label{fig:spectral_artificial_data}Artificial data and forward model for the spectral deconvolution model at (a) $K=3$, (b) $10$, and (c) $30$. Blue dots: observed data; red solid lines: forward model with the true parameters; gray dotted lines: individual peak components.}
\end{figure}

For REMC(CPU), the number of MCMC steps (including burn-in) was varied using a base unit of 200 multiplied by $\{1, 3, 10, 30, 100, 300\}$ for $K{=}3$ and by $\{1, 3, 10, 30\}$ for $K{=}10$ and $K{=}30$; the burn-in ratio was 50\%.
For SMCS(GPU), the number of MCMC steps per temperature level was $n=10$ for $K{=}3$ and $10$, and $n=100$ for $K{=}30$.
The particle count was varied by multiplying a base unit of 100 by $\{3, 10, 30, 100, 300, 1000, 3000, 10\,000\}$, with an additional factor of $30\,000$ for $K{=}30$.
Each condition was repeated independently 100 times for $K{=}3$ and $K{=}30$; for $K{=}10$, REMC and SMCS were repeated 10 and 100 times, respectively.
For $K{=}30$ at the smallest REMC(CPU) sample size (200 steps), the free-energy estimate diverged in 6 out of 100 runs; these runs were excluded from the analysis.

\paragraph{Free-energy comparison.}

\begin{figure}
  \includegraphics[width=\columnwidth]{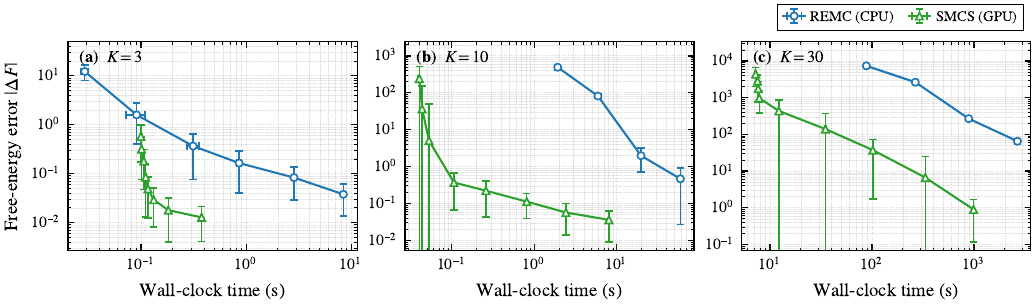}
  \caption{\label{fig:spectral_fe_error}Free-energy error $|\Delta F|$ versus computation time for the spectral deconvolution model at (a) $K=3$, (b) $10$, and (c) $30$. Blue circles: REMC(CPU); green triangles: SMCS(GPU). For $K{=}30$ at the minimum REMC(CPU) sample size, divergent runs (6/100) were excluded.}
\end{figure}

Figure~\ref{fig:spectral_fe_error} plots the free-energy error $|\Delta F|$ against computation time for each $K$.
The reference value is the trial-averaged SMCS(GPU) result at the largest particle count.

At every $K$, SMCS(GPU) requires substantially less computation time than REMC(CPU) to reach a comparable free-energy accuracy.
The speedup factors are approximately $70\times$ for $K{=}3$, $591\times$ for $K{=}10$, and $41\times$ for $K{=}30$.
The particularly large speedup at $K{=}10$ and the reduced speedup at $K{=}30$---where the increased GPU cost due to the larger number of MCMC steps $n$ becomes significant---are discussed further in Sec.~\ref{sec:discussion}.
Table~\ref{tab:spectral_speedup} summarizes the computation times and speedup ratios.

\begin{table}
  \caption{\label{tab:spectral_speedup}Speedup of SMCS(GPU) over REMC(CPU) for the spectral deconvolution model at each peak count $K$. Computed from the wall-clock time at which SMCS(GPU) attains the same $|\Delta F|$ as the largest REMC(CPU) run.}
  \begin{ruledtabular}
  \begin{tabular}{cccc}
    $K$ & REMC time (s) & SMCS time (s) & Speedup \\
    \hline
    3   & 8.4     & 0.12  & $70\times$  \\
    10  & 59.1    & 0.10  & $591\times$ \\
    30  & 2663.1  & 65.19 & $41\times$  \\
  \end{tabular}
  \end{ruledtabular}
\end{table}

The signed error $F - F_{\mathrm{ref}}$ is plotted against computation time in Fig.~\ref{fig:spectral_fe_value} (Appendix~\ref{sec:appendix_fe_convergence}).
Both methods converge to zero error with increasing sample size, confirming that they sample from the same target distribution.

\paragraph{Comparison of posterior credible intervals.}

\begin{figure}
  \includegraphics[width=\columnwidth]{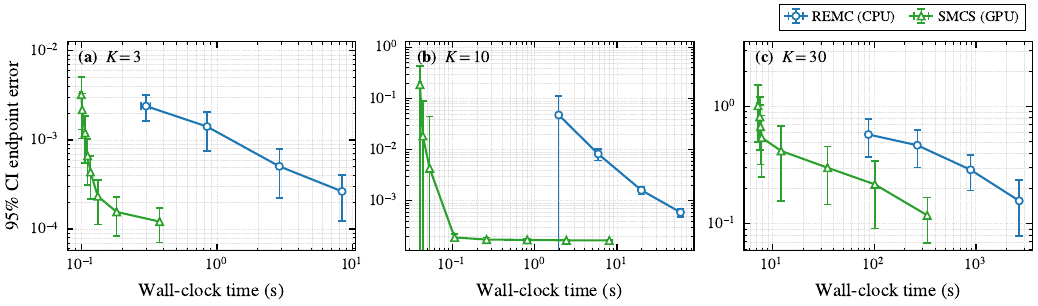}
  \caption{\label{fig:spectral_param_error}Sum of 95\% credible-interval endpoint errors for the peak center $\mu$, averaged over all $K$ peaks, versus computation time (spectral deconvolution model). Results for (a) $K=3$, (b) $10$, and (c) $30$ are shown. Blue circles: REMC(CPU); green triangles: SMCS(GPU). For $K{=}30$ at the minimum REMC(CPU) sample size, divergent runs (6/100) were excluded.}
\end{figure}

Figure~\ref{fig:spectral_param_error} shows the convergence of the 95\% credible interval for the peak center position $\mu$, plotting the summed endpoint error averaged over all $K$ peaks against computation time.
The baseline is the trial-averaged endpoint value at the largest SMCS(GPU) particle count.
As with the free energy, SMCS(GPU) converges to the baseline accuracy in less wall-clock time than REMC(CPU) at every $K$.

\subsection{\label{sec:real_data}Real data}

Although artificial data allow comparison with known ground truth, the analysis of real measurements is indispensable in practice.
Real data involve factors absent from synthetic benchmarks---measurement noise with nontrivial statistical properties, complex background shapes, and potential mismatch between the model and the actual spectrum---making it essential to verify that the proposed method performs reliably under realistic conditions.
We compare SMCS(GPU) and REMC(CPU) on measured XRD and XPS spectra.

\subsubsection{\label{sec:xrd_real}XRD data}
\paragraph{Experimental setup.}

The real XRD data consist of a powder diffraction pattern from a 1:1 mass-ratio mixture of TiO$_2$ rutile and anatase.
Measurements were carried out at 40\,kV/50\,mA using a Cu~K$\beta$-filtered one-dimensional scan mode with a step width of $0.02^{\circ}$, a scan speed of $4.00\,{}^{\circ}/\text{min}$, a scan range of $5^{\circ}$--$60^{\circ}$ ($2\theta/\theta$), and a HyPix-3000 detector in horizontal orientation.
Only data points with $2\theta \geq 20^{\circ}$ ($N=2001$) were used in the analysis.
Full details of the sample and instrumentation are given in Appendix~\ref{sec:appendix_xrd}.

\begin{figure}
  \includegraphics[width=\columnwidth]{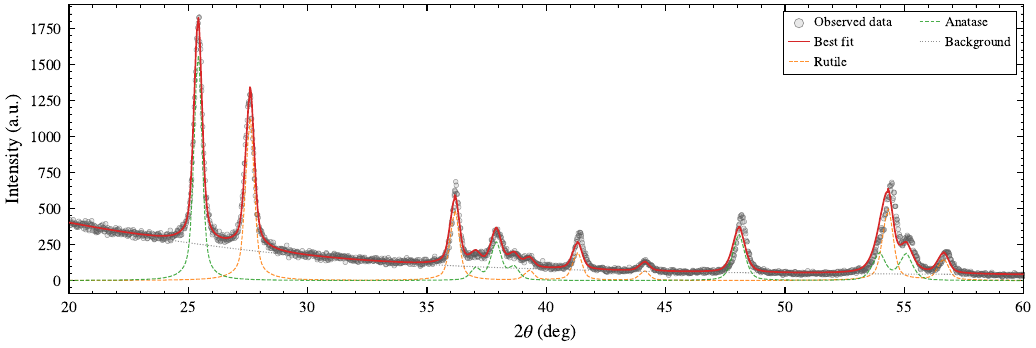}
  \caption{\label{fig:real_xrd_fit}Measured XRD pattern of the TiO$_2$ mixed-powder sample and the fitting result. The best fit (blue line) to the observed data (black line) is shown together with the rutile (orange dashed), anatase (green dashed), and background (red dotted) contributions.}
\end{figure}

The forward model is the same pseudo-Voigt function used in Sec.~\ref{sec:xrd_artificial}, but here a two-phase model (rutile and anatase; $2\times 9 + 4 = 22$ parameters) is employed, omitting the brookite phase.
Figure~\ref{fig:real_xrd_fit} shows the measured spectrum together with the best fit from REMC(CPU) (the minimum-energy sample), decomposed into the contributions of each phase and the background.

For REMC(CPU), the total number of MCMC steps (including burn-in) was varied by multiplying a base unit of 100 by $\{1, 3, 10, 30, 100, 300, 1000\}$; the burn-in ratio was 50\% and the number of replicas was $L=38$.
For SMCS(GPU), the number of MCMC steps per temperature level was $n=100$, and the particle count was varied by multiplying a base unit of 100 by $\{1, 3, 10, 30, 100, 300, 1000, 3000, 10\,000\}$.
Each condition was repeated independently 100 times.

\paragraph{Free-energy comparison.}

\begin{figure}
  \includegraphics[width=0.48\columnwidth]{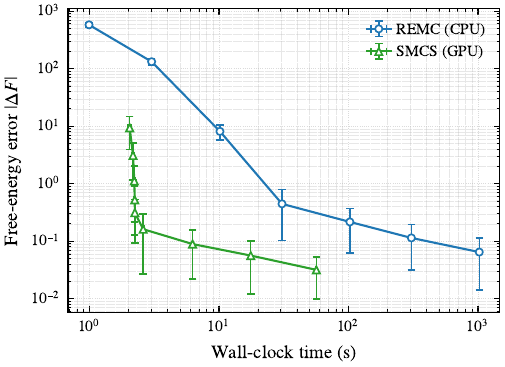}
  \caption{\label{fig:real_xrd_fe_error}Free-energy error $|\Delta F|$ versus computation time for the XRD real data. Blue circles: REMC(CPU); green triangles: SMCS(GPU).}
\end{figure}

Figure~\ref{fig:real_xrd_fe_error} shows the free-energy error $|\Delta F|$ versus computation time.
Because the true free energy is unknown, the trial-averaged SMCS(GPU) value at $10^6$ particles serves as the reference.
The speedup required for SMCS(GPU) to match the accuracy of the largest REMC(CPU) run is approximately $79\times$ (Table~\ref{tab:real_xrd_speedup}).

\begin{table}
  \caption{\label{tab:real_xrd_speedup}Speedup of SMCS(GPU) over REMC(CPU) for XRD real data.}
  \begin{ruledtabular}
  \begin{tabular}{ccc}
    REMC time (s) & SMCS time (s) & Speedup \\
    \hline
    1023.7 & 12.87 & $79\times$ \\
  \end{tabular}
  \end{ruledtabular}
\end{table}

The signed error $F - F_{\mathrm{ref}}$ is plotted against computation time in Fig.~\ref{fig:real_xrd_fe_value} (Appendix~\ref{sec:appendix_fe_convergence}).
Both methods converge to zero error with increasing sample size, confirming that they sample from the same target distribution even on real data.

\paragraph{Comparison of posterior credible intervals.}

\begin{figure}
  \includegraphics[width=0.48\columnwidth]{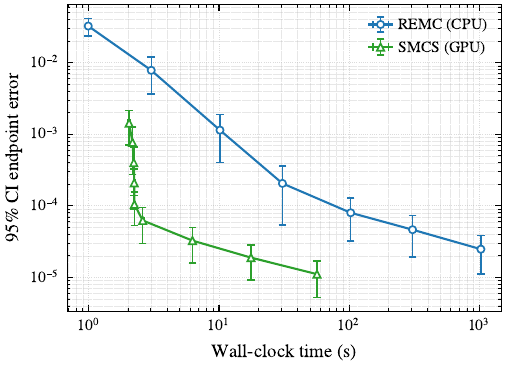}
  \caption{\label{fig:real_xrd_param_error}Sum of 95\% credible-interval endpoint errors for the rutile $2\theta$ shift parameter versus computation time (XRD real data). Blue circles: REMC(CPU); green triangles: SMCS(GPU).}
\end{figure}

Figure~\ref{fig:real_xrd_param_error} shows the summed 95\% credible-interval endpoint error for the rutile $2\theta$ shift parameter versus computation time.
The baseline is the trial-averaged endpoint value at the largest SMCS(GPU) particle count.
As observed for the artificial data, SMCS(GPU) reaches the same convergence level in less wall-clock time than REMC(CPU).

\subsubsection{\label{sec:xps_real}XPS data}
\paragraph{Experimental setup.}

We use the hard X-ray photoelectron spectroscopy (HAXPES) spectrum of Ni$_3$Al$_2$O$_3$ published by Longo \textit{et al.}~\cite{Longo_DeepCoreLevelHAXPES_2023} as the experimental dataset (840 data points covering binding energies of approximately 845--887\,eV).

The model consists of pseudo-Voigt peak functions superimposed on a Shirley background (see Appendix~\ref{sec:appendix_xps} for the full formulation and prior distributions).
Each peak has four parameters: amplitude $A_k$, position $\mu_k$, half-width at half-maximum $\sigma_k$, and Gaussian--Lorentzian mixing ratio $\eta_k$.
The background is parameterized by two Shirley-function endpoint intensities $a$ and $b$.
Model selection is performed by comparing three candidate peak counts, $K=6$, $7$, and $8$, based on the Bayesian free energy.
For $K{=}7$, the total number of estimated parameters is $7 \times 4 + 2 = 30$.

\begin{figure}
  \includegraphics[width=0.66\columnwidth]{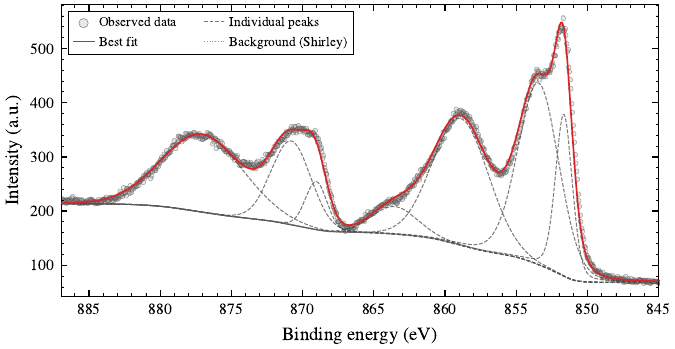}
  \caption{\label{fig:real_xps_fit}HAXPES measured spectrum of Ni$_3$Al$_2$O$_3$ and the fitting result for the $K{=}7$ model. The best fit (blue line) to the observed data (black line) is shown together with each pseudo-Voigt peak component (gray dotted) and the Shirley background (red dotted).}
\end{figure}

Figure~\ref{fig:real_xps_fit} displays the measured spectrum together with the REMC(CPU) best fit for the $K{=}7$ model, decomposed into individual pseudo-Voigt peak components and the Shirley background.

For REMC(CPU), the total MCMC steps (including burn-in) were varied by multiplying a base unit of 200 by $\{1, 3, 10, 30, 100, 300, 1000\}$, with a burn-in ratio of 50\%.
For SMCS(GPU), the number of MCMC steps per temperature level was $n=100$, and the particle count was varied by multiplying a base unit of 200 by $\{3, 10, 30, 100, 300, 1000, 3000, 10\,000\}$.
Each condition was repeated independently 100 times.

\paragraph{Free-energy comparison.}

\begin{figure}
  \includegraphics[width=\columnwidth]{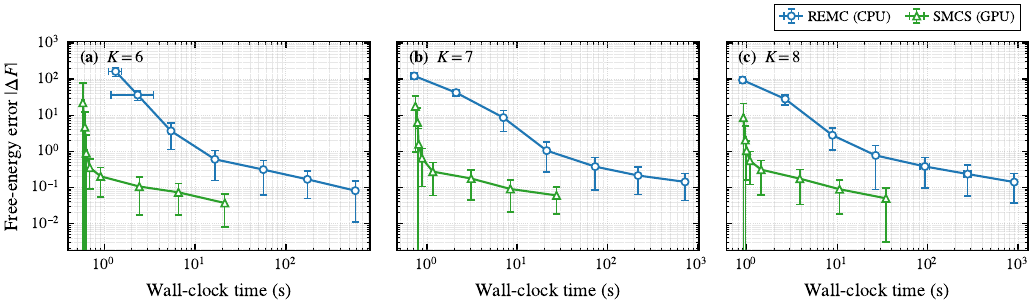}
  \caption{\label{fig:real_xps_fe_error}Free-energy error $|\Delta F|$ versus computation time for XPS real data at (a) $K=6$, (b) $7$, and (c) $8$. Blue circles: REMC(CPU); green triangles: SMCS(GPU).}
\end{figure}

Figure~\ref{fig:real_xps_fe_error} shows the free-energy error $|\Delta F|$ versus computation time for $K=6$, $7$, and $8$.
The reference is the trial-averaged SMCS(GPU) value at the largest particle count.
The speedup factors are $118\times$ ($K{=}6$), $172\times$ ($K{=}7$), and $169\times$ ($K{=}8$); see Table~\ref{tab:real_xps_speedup}.

\begin{table}
  \caption{\label{tab:real_xps_speedup}Speedup of SMCS(GPU) over REMC(CPU) for XPS real data at each peak count $K$.}
  \begin{ruledtabular}
  \begin{tabular}{cccc}
    $K$ & REMC time (s) & SMCS time (s) & Speedup \\
    \hline
    6  & 580.9 & 4.93 & $118\times$ \\
    7  & 732.9 & 4.26 & $172\times$ \\
    8  & 905.6 & 5.36 & $169\times$ \\
  \end{tabular}
  \end{ruledtabular}
\end{table}

The signed error $F - F_{\mathrm{ref}}$ is plotted against computation time in Fig.~\ref{fig:real_xps_fe_value} (Appendix~\ref{sec:appendix_fe_convergence}).
For all three models, both methods converge to zero error.
Among the converged free-energy values, $K{=}7$ gives the smallest value ($F \approx 3232.0$), followed closely by $K{=}8$ ($F \approx 3232.5$), with both well below $K{=}6$ ($F \approx 3256.0$).
The Bayesian model selection criterion therefore identifies $K{=}7$ as the optimal peak count.

\paragraph{Model-selection accuracy at comparable computation times.}

\begin{figure}
  \includegraphics[width=0.48\columnwidth]{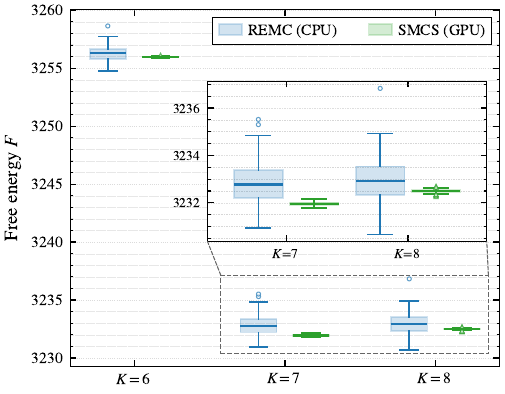}
  \caption{\label{fig:real_xps_fe_boxplot}Box plots of the Bayesian free energy for the XPS real data ($K=6, 7, 8$). REMC(CPU) at 6000 steps (blue) and SMCS(GPU) at $2\times10^6$ particles (green) are compared at similar computation times. Each box summarizes 100 independent runs. The inset magnifies the $K{=}7$ and $K{=}8$ region, highlighting the markedly smaller variability of SMCS(GPU).}
\end{figure}

To compare the two methods at similar wall-clock times, we select REMC(CPU) at 6000 steps (approximately 16--27\,s depending on $K$) and SMCS(GPU) at $2\times 10^6$ particles (approximately 21--35\,s).
Figure~\ref{fig:real_xps_fe_boxplot} presents box plots of the free-energy distributions over 100 independent runs for $K=6$, $7$, and $8$.
The REMC(CPU) free-energy standard deviation ranges from 0.68 to 0.96, exceeding the gap between $K{=}7$ and $K{=}8$ ($\Delta F \approx 0.5$); consequently, the model-selection outcome can fluctuate from run to run.
By contrast, SMCS(GPU) reduces the standard deviation to 0.05--0.07---more than tenfold smaller---allowing the $K{=}7$/$K{=}8$ difference to be resolved reliably.
In other words, at comparable computation times, SMCS(GPU) estimates the free energy with substantially higher precision than REMC(CPU), enabling stable model selection even when the free-energy differences between competing models are small.

\paragraph{Comparison of posterior credible intervals.}

\begin{figure}
  \includegraphics[width=0.48\columnwidth]{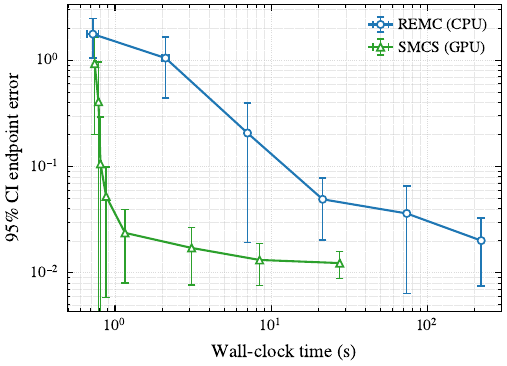}
  \caption{\label{fig:real_xps_param_error}Sum of 95\% credible-interval endpoint errors for the peak position $\mu$ versus computation time for XPS real data ($K{=}7$ model). The error, averaged over all seven peaks, is shown on the vertical axis. Blue circles: REMC(CPU); green triangles: SMCS(GPU).}
\end{figure}

Figure~\ref{fig:real_xps_param_error} shows the summed 95\% credible-interval endpoint error for the peak position $\mu$ in the $K{=}7$ model, averaged over all seven peaks, as a function of computation time.
The baseline is the trial-averaged endpoint value at the largest SMCS(GPU) particle count.
As in the XRD case, SMCS(GPU) converges to the same accuracy in less wall-clock time than REMC(CPU).

\section{\label{sec:discussion}Discussion}

We now discuss the trends in the speedup ratios obtained across all experiments (Tables~\ref{tab:xrd_speedup}--\ref{tab:real_xps_speedup}) and the factors that govern the efficiency of GPU parallelization.

\paragraph{Dependence on data size $N$ (XRD artificial data).}

For the XRD artificial data, the speedup ratios at $N=1000$, $5000$, and $10\,000$ are approximately $316\times$, $547\times$, and $532\times$, respectively (Table~\ref{tab:xrd_speedup}).
In REMC(CPU), parallelism is limited to the number of replicas $L$ (on the order of tens), and the likelihood evaluation within each replica scales roughly linearly with $N$.
SMCS(GPU), by contrast, parallelizes over both particles and data points, so that the computation time grows more slowly with $N$ and the speedup ratio increases (Fig.~\ref{fig:xrd_fe_error}).
Between $N{=}5000$ and $N{=}10\,000$, however, the speedup plateaus: because SMCS(GPU) parallelizes simultaneously over $N$ and the particle count, the total degree of parallelism already exceeds the GPU's available cores at $N{=}5000$, saturating the benefit of further increases in $N$.
Even so, a speedup exceeding $500\times$ is maintained at $N{=}10\,000$, underscoring the effectiveness of GPU parallelization for large-scale data analysis.

\paragraph{Dependence on peak count $K$ (spectral deconvolution artificial data).}

For the spectral deconvolution model, the speedup ratios are approximately $70\times$ ($K{=}3$), $591\times$ ($K{=}10$), and $41\times$ ($K{=}30$)---a nonmonotonic trend (Table~\ref{tab:spectral_speedup}).
This behavior can be understood as a competition among three factors: (i) the computational cost per likelihood evaluation, which scales roughly as $O(NK)$; (ii) fixed overhead that is non-negligible relative to the parallelizable workload (resampling, random-number generation, kernel launches); and (iii) the difference in sampling efficiency between the SMCS and REMC as the parameter dimension increases.
At $K{=}3$, both the data size ($N{=}300$) and the parallelizable workload are small, so fixed overhead dominates and limits the speedup.
At $K{=}10$ ($N{=}1000$), the cost of evaluating the peak superposition is substantial while the SMCS still mixes efficiently, yielding the highest speedup.
At $K{=}30$, the parameter dimension reaches $3K{=}90$, and SMCS mixing degrades; the sampling efficiency per unit computation falls relative to REMC(CPU), narrowing the performance gap despite GPU parallelization.
Thus, while increasing $K$ provides more work to parallelize, the accompanying growth in dimension erodes the SMCS's sampling advantage, producing the observed nonmonotonic dependence.

\paragraph{Speedup on real data.}

On real data, speedups of approximately $79\times$ (XRD, Table~\ref{tab:real_xrd_speedup}) and $118$--$172\times$ (XPS, Table~\ref{tab:real_xps_speedup}) are obtained.
Although smaller than the peak values on artificial data ($>500\times$), these ratios confirm that GPU parallelization remains effective under realistic conditions.
Two factors contribute to the reduced speedup.
First, the discrepancy between the theoretical model and the actual observations---absent in artificial data---roughens the energy landscape and impairs SMCS mixing, lowering its sampling efficiency relative to REMC.
Second, the forward model for real data involves transcendental-function evaluations (pseudo-Voigt profiles, angle-dependent widths, Shirley backgrounds) that reduce GPU throughput compared with the pure-Gaussian model used for artificial data.
Nonetheless, SMCS(GPU) completes Bayesian model selection on real data in tens of seconds per model---a practically significant result.
For instance, on the XPS real data, the Bayesian free energy for each of the three candidate peak counts ($K{=}6, 7, 8$) can be computed in tens of seconds (Table~\ref{tab:real_xps_speedup}), enabling automated end-to-end analysis from peak-count selection through parameter estimation.

\paragraph{Summary.}

For problems with large data and moderate model complexity, GPU parallelization achieves speedups exceeding $500\times$.
When the parameter dimension is high ($K{=}30$) or the energy landscape is complex (real data), speedups range from $40\times$ to $170\times$, limited primarily by the SMCS's reduced mixing efficiency.
Improving the mixing performance of the SMCS for complex energy landscapes and high-dimensional parameter spaces remains an important challenge for extending the method to more demanding problem settings.

\section{\label{sec:conclusion}Conclusion}

We have proposed a GPU-accelerated implementation of the sequential Monte Carlo sampler (SMCS) for Bayesian spectral analysis and conducted a systematic performance comparison with CPU-parallelized replica exchange Monte Carlo (REMC).

On artificial data for XRD and spectral-deconvolution models, the GPU-parallelized SMCS achieves speedups exceeding $500\times$ over REMC(CPU), as measured by convergence of the Bayesian free energy.
The speedup ratio increases with data size $N$, remaining above $500\times$ even at $N{=}10\,000$.
With respect to model dimensionality, the largest speedup of $591\times$ is attained at $K{=}10$ peaks, whereas at $K{=}30$ it falls to $41\times$---a reduction attributed primarily to degraded SMCS mixing in the resulting 90-dimensional parameter space.
The 95\% posterior credible intervals likewise converge faster under SMCS(GPU) than under REMC(CPU).

On measured XRD and XPS data, speedups of $79\times$ to $172\times$ are obtained.
The smaller ratios compared to artificial data are ascribed to the increased complexity of the energy landscape arising from discrepancies between the theoretical model and real observations.
Importantly, SMCS(GPU) completes Bayesian model selection in tens of seconds per model on real data, demonstrating that fully automated analysis---from model selection through parameter estimation---is now practical.
As measurement techniques such as microscopic spectroscopy and in-situ methods continue to produce rapidly growing volumes of spectral data, the proposed approach provides a foundational technology for automating Bayesian model selection and parameter estimation, contributing to the advancement of measurement-data analysis in materials science.

A key direction for future work is to improve the mixing efficiency of the SMCS in high-dimensional and complex-landscape problems.
As demonstrated in this study, the sampling-efficiency gap between the SMCS and REMC widens with problem difficulty, calling for the development of methods that balance sampling efficiency with parallelization efficiency---for example, through Hamiltonian Monte Carlo kernels, gradient-informed proposals, or adaptive tempering schedules that maintain both rapid mixing and massive parallelism.

\begin{acknowledgments}
This work was supported by ISUZU MOTORS LIMITED through funding for the purchase of computational equipment and personnel costs. The funder had no role in the study design, data collection, analysis, or in manuscript preparation.
\end{acknowledgments}

\appendix

\section{\label{sec:computational_env}Computational environment}

The numerical experiments in this study were performed on the hardware and software listed in Table~\ref{tab:computational_env}.
REMC(CPU) was implemented in C++ (g++ 13.3.0) with OpenMP parallelization; SMCS(GPU) was implemented in CUDA 12.8.

\begin{table}
  \caption{\label{tab:computational_env}Hardware and software used in the numerical experiments.}
  \begin{ruledtabular}
  \begin{tabular}{ll}
    Component & Specification \\
    \hline
    OS & Ubuntu 24.04.2 LTS \\
    CPU & AMD Ryzen 9 9950X (16 cores / 32 threads) \\
    GPU & NVIDIA GeForce RTX 5090 \\
    Main memory & 128\,GB \\
    \hline
    REMC(CPU) implementation & C++ (g++ 13.3.0), OpenMP \\
    SMCS(GPU) implementation & CUDA 12.8 \\
  \end{tabular}
  \end{ruledtabular}
\end{table}

\section{\label{sec:appendix_fe_convergence}Convergence of free-energy estimates}

\begin{figure}
  \includegraphics[width=\columnwidth]{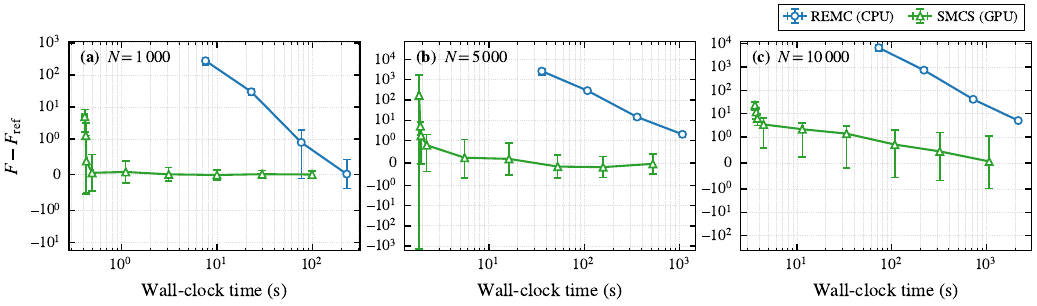}
  \caption{\label{fig:xrd_fe_value}Free-energy estimation error $F - F_{\mathrm{ref}}$ versus computation time for XRD artificial data: (a) $N=1000$, (b) $5000$, (c) $10\,000$.}
\end{figure}

\begin{figure}
  \includegraphics[width=\columnwidth]{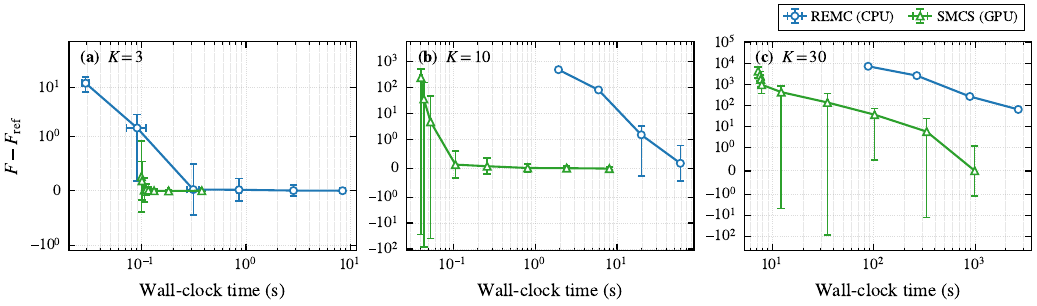}
  \caption{\label{fig:spectral_fe_value}Free-energy estimation error $F - F_{\mathrm{ref}}$ versus computation time for the spectral deconvolution model: (a) $K=3$, (b) $10$, (c) $30$. For $K{=}30$ at the minimum REMC(CPU) sample size, divergent runs (6/100) were excluded.}
\end{figure}

\begin{figure}
  \includegraphics[width=0.48\columnwidth]{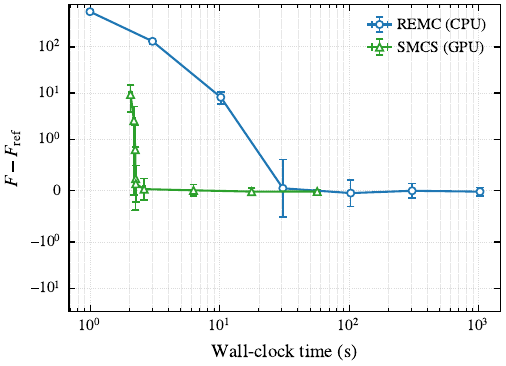}
  \caption{\label{fig:real_xrd_fe_value}Free-energy estimation error $F - F_{\mathrm{ref}}$ versus computation time for XRD real data.}
\end{figure}

\begin{figure}
  \includegraphics[width=\columnwidth]{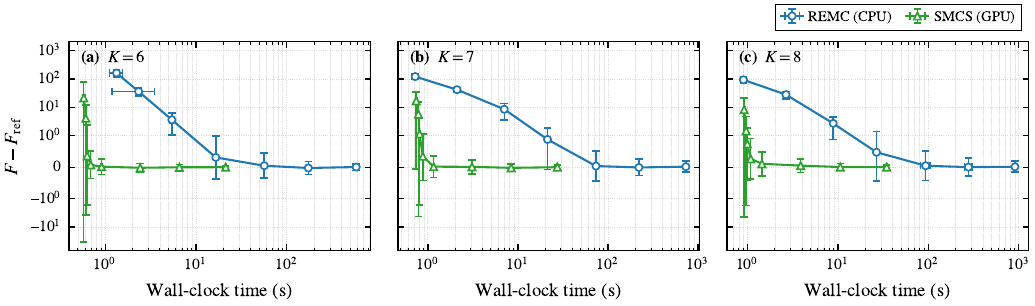}
  \caption{\label{fig:real_xps_fe_value}Free-energy estimation error $F - F_{\mathrm{ref}}$ versus computation time for XPS real data: (a) $K=6$, (b) $7$, (c) $8$.}
\end{figure}

Figures~\ref{fig:xrd_fe_value}--\ref{fig:real_xps_fe_value} display the signed free-energy estimation error $F - F_{\mathrm{ref}}$ as a function of computation time for every experiment reported in Sec.~\ref{sec:experiments}.
Here $F_{\mathrm{ref}}$ denotes the trial average at the largest SMCS(GPU) particle count; the vertical axis uses a symmetric-log scale, so values near zero are displayed linearly.
In every case, both REMC(CPU) and SMCS(GPU) converge to $F - F_{\mathrm{ref}} \to 0$ with increasing sample size, confirming that both samplers target the same distribution.

\section{\label{sec:appendix_artificial_model}Forward models for artificial data}

\subsection{Spectral deconvolution model}

The artificial data for the spectral deconvolution experiments (Sec.~\ref{sec:spectral_artificial}) were generated by adding Gaussian noise to a superposition of Gaussian peaks.
The forward model is
\begin{equation}
  f(x;\theta) = \sum_{k=1}^{K} A_k \exp\!\left(-\frac{b_k}{2}(x - \mu_k)^2\right).
  \label{eq:spectral_forward_model}
\end{equation}
Each peak has three parameters: amplitude $A_k$, center $\mu_k$, and width $b_k$.
Observed data $y_i$ were generated as $y_i = f(x_i;\theta) + \varepsilon_i$ with $\varepsilon_i \sim \mathcal{N}(0,\sigma^2)$.
The noise standard deviation $\sigma$ was treated as known and held fixed; the corresponding negative log-likelihood (energy function) is
\begin{equation}
  E(\theta) = \frac{1}{2N\sigma^2}\sum_{i=1}^N (y_i - f(x_i;\theta))^2.
\end{equation}
Table~\ref{tab:spectral_true_params} lists the experimental conditions used for data generation, and Tables~\ref{tab:spectral_true_K3}--\ref{tab:spectral_true_K30} give the true parameter values for each peak count.
For $K{=}3$, the parameters were set manually.
For $K{=}10$ and $K{=}30$, they were drawn from pseudorandom distributions: $A_k \sim \mathrm{U}(0.5, 1.5)$, $b_k \sim \mathrm{U}(100, 200)$, and $\mu_k \sim \mathcal{N}(0.5+k, \sigma_\mu^2)$ with $\sigma_\mu = 0.4$ ($K{=}10$) or $\sigma_\mu = 0.3$ ($K{=}30$).

\begin{table}
  \caption{\label{tab:spectral_true_params}Experimental conditions for generating the spectral deconvolution artificial data.}
  \begin{ruledtabular}
  \begin{tabular}{cccc}
    $K$ & $N$ & $x$ range & $\sigma$ \\
    \hline
    3  & 300  & $[0, 3]$  & 0.1  \\
    10 & 1000 & $[0, 10]$ & 0.1  \\
    30 & 3000 & $[0, 30]$ & 0.05 \\
  \end{tabular}
  \end{ruledtabular}
\end{table}

\begin{table}
  \caption{\label{tab:spectral_true_K3}True parameter values for $K=3$.}
  \begin{ruledtabular}
  \begin{tabular}{cccc}
    $k$ & $A_k$ & $\mu_k$ & $b_k$ \\
    \hline
    1 & 0.587 & 1.210 & 95.689  \\
    2 & 1.522 & 1.455 & 146.837 \\
    3 & 1.183 & 1.703 & 164.469 \\
  \end{tabular}
  \end{ruledtabular}
\end{table}

\begin{table}
  \caption{\label{tab:spectral_true_K10}True parameter values for $K=10$.}
  \begin{ruledtabular}
  \begin{tabular}{cccc}
    $k$ & $A_k$ & $\mu_k$ & $b_k$ \\
    \hline
    1  & 1.049 & 0.558  & 114.335 \\
    2  & 1.215 & 2.082  & 194.467 \\
    3  & 1.103 & 2.804  & 152.185 \\
    4  & 1.045 & 3.549  & 141.466 \\
    5  & 0.924 & 4.678  & 126.456 \\
    6  & 1.146 & 5.633  & 177.423 \\
    7  & 0.938 & 7.098  & 145.615 \\
    8  & 1.392 & 7.418  & 156.843 \\
    9  & 1.464 & 8.625  & 101.879 \\
    10 & 0.883 & 9.158  & 161.764 \\
  \end{tabular}
  \end{ruledtabular}
\end{table}

\begin{table}
  \caption{\label{tab:spectral_true_K30}True parameter values for $K=30$.}
  \begin{ruledtabular}
  \begin{tabular}{cccc}
    $k$ & $A_k$ & $\mu_k$ & $b_k$ \\
    \hline
    1  & 0.576 & 0.486  & 105.313 \\
    2  & 1.280 & 1.065  & 130.885 \\
    3  & 0.938 & 2.378  & 159.259 \\
    4  & 1.223 & 2.814  & 123.512 \\
    5  & 1.478 & 4.815  & 196.497 \\
    6  & 1.038 & 5.375  & 194.505 \\
    7  & 1.001 & 6.277  & 184.840 \\
    8  & 0.572 & 7.822  & 147.232 \\
    9  & 0.768 & 8.005  & 184.148 \\
    10 & 1.000 & 9.661  & 113.111 \\
    11 & 1.179 & 9.881  & 130.873 \\
    12 & 1.304 & 11.301 & 146.300 \\
    13 & 0.881 & 12.139 & 174.185 \\
    14 & 0.566 & 13.939 & 148.583 \\
    15 & 0.788 & 15.030 & 113.688 \\
    16 & 1.410 & 15.401 & 134.354 \\
    17 & 0.713 & 16.752 & 132.443 \\
    18 & 0.952 & 17.446 & 130.042 \\
    19 & 1.431 & 18.670 & 116.550 \\
    20 & 0.525 & 19.274 & 141.490 \\
    21 & 1.101 & 19.987 & 144.812 \\
    22 & 1.450 & 20.959 & 177.490 \\
    23 & 0.730 & 22.615 & 179.639 \\
    24 & 1.048 & 24.174 & 152.239 \\
    25 & 1.409 & 24.581 & 146.063 \\
    26 & 0.633 & 25.343 & 177.821 \\
    27 & 1.023 & 27.074 & 188.729 \\
    28 & 1.250 & 27.571 & 167.492 \\
    29 & 1.169 & 28.530 & 180.048 \\
    30 & 0.968 & 29.576 & 193.911 \\
  \end{tabular}
  \end{ruledtabular}
\end{table}

Table~\ref{tab:spectral_priors} lists the prior distributions.
Gamma priors enforce the positivity of $A_k$ and $b_k$.
For $\mu_k$, a normal prior is used at $K{=}3$, whereas uniform priors spanning the $x$ range are used for $K{=}10$ and $K{=}30$.

\begin{table}
  \caption{\label{tab:spectral_priors}Prior distributions for the spectral deconvolution model. $\mathrm{Ga}(\alpha,\beta)$: gamma (shape $\alpha$, rate $\beta$); $\mathcal{N}(\mu,\sigma^2)$: normal; $\mathrm{U}(a,b)$: uniform.}
  \begin{ruledtabular}
  \begin{tabular}{ccccc}
    Parameter & Description & $K=3$ & $K=10$ & $K=30$ \\
    \hline
    $A_k$    & Amplitude & $\mathrm{Ga}(5, 5)$     & $\mathrm{Ga}(5, 5)$     & $\mathrm{Ga}(5, 5)$     \\
    $\mu_k$  & Center    & $\mathcal{N}(1.5, 0.2)$ & $\mathrm{U}(0, 10)$     & $\mathrm{U}(0, 30)$     \\
    $b_k$    & Width     & $\mathrm{Ga}(5, 0.04)$  & $\mathrm{Ga}(5, 0.04)$  & $\mathrm{Ga}(5, 0.04)$  \\
  \end{tabular}
  \end{ruledtabular}
\end{table}

\subsection{XRD model}

The XRD artificial data (Sec.~\ref{sec:xrd_artificial}) were generated using the pseudo-Voigt forward model described in Appendix~\ref{sec:appendix_xrd} [Eq.~\eqref{eq:xrd_forward_model}].
Data sizes of $N=1000$, $5000$, and $10\,000$ were used.
The artificial spectra were constructed from three crystalline phases (rutile, anatase, brookite) and a pseudo-Voigt background.
Tables~\ref{tab:xrd_true_phase} and~\ref{tab:xrd_true_bg} list the true parameter values used for data generation.

\begin{table*}
  \caption{\label{tab:xrd_true_phase}True crystalline-phase parameters for the XRD artificial data. $A_k$: peak intensity scale; $\Delta 2\theta_k$: peak position shift (degrees); $\alpha_k$: asymmetry factor; $r_k$: mixing ratio; $u_k$, $v_k$, $w_k$: Caglioti parameters; $s_k$, $t_k$: Lorentzian width parameters.}
  \begin{ruledtabular}
  \begin{tabular}{lccccccccc}
    Phase & $A_k$ & $\Delta 2\theta_k$ & $\alpha_k$ & $r_k$ & $u_k$ & $v_k$ & $w_k$ & $s_k$ & $t_k$ \\
    \hline
    Rutile    & 10000 & 0.035 & 0.6 & 0.50 & 0.03 & 0.03 & 0.06 & 0.06 & 0.03 \\
    Anatase   & 3500  & 0.055 & 0.9 & 0.65 & 0.1  & 0.1  & 0.2  & 0.2  & 0.1  \\
    Brookite  & 1000  & 0.04  & 1.0 & 0.75 & 0.1  & 0.1  & 0.2  & 0.2  & 0.1  \\
  \end{tabular}
  \end{ruledtabular}
\end{table*}

\begin{table}
  \caption{\label{tab:xrd_true_bg}True background parameters for the XRD artificial data.}
  \begin{ruledtabular}
  \begin{tabular}{cccc}
    $a$ & $\sigma_{\mathrm{bg}}$ & $r_{\mathrm{bg}}$ & $b$ \\
    \hline
    60000 & 10 & 0.0 & 100 \\
  \end{tabular}
  \end{ruledtabular}
\end{table}

\section{\label{sec:appendix_xrd}XRD data and model formulation}

X-ray diffraction (XRD) analyzes crystal structures by irradiating a sample with X-rays and recording the diffraction peaks.
Let the observed data be $D=\{(x_i,y_i)\}_{i=1}^{N}$, where $x_i$ is the diffraction angle $2\theta$ (degrees) and $y_i$ is the corrected intensity.
Following the powder-diffraction profile model of Murakami \textit{et al.}~\cite{murakami2024bayesian}, we adopt pseudo-Voigt basis functions.
The forward model is
\begin{equation}
\begin{aligned}
  f(x, \Theta)
  &= \sum_{k=1}^{K} A_k\,V\!\left(x;\mu_k,\Sigma_k(x),\Omega_k(x),r_k\right)
  + B(x;\Theta_B),
\end{aligned}
\label{eq:xrd_forward_model}
\end{equation}
where $A_k>0$ is the peak intensity, $\mu_k$ is the peak position, and $r_k\in[0,1]$ is the Gaussian--Lorentzian mixing ratio.
The pseudo-Voigt function $V$ is
\begin{equation}
  V(x;\rho,\Sigma,\Omega,r)=(1-r)\,G(x;\rho,\Sigma)+r\,L(x;\rho,\Omega),
  \label{eq:xrd_pvoigt}
\end{equation}
where the Gaussian and Lorentzian components (with normalization constants absorbed into $A_k$) are given by
\begin{align}
  G(x;\rho,\Gamma_G) &= \exp\!\left[-4\ln 2\left(\frac{x-\rho}{\Gamma_G}\right)^2\right], \label{eq:xrd_gauss}\\
  L(x;\rho,\Gamma_L) &= \frac{1}{1+4\left(\frac{x-\rho}{\Gamma_L}\right)^2}. \label{eq:xrd_lorentz}
\end{align}
The angle-dependent width parameters $\Sigma_k(x)$ (Gaussian) and $\Omega_k(x)$ (Lorentzian) are given by
\begin{align}
  \Sigma_k(x)
  &= A_{\mathrm{asym}}(x;\alpha_k,\mu_k)
  \sqrt{u_k\tan^2\!\left(\frac{x}{2}\right)-v_k\tan\!\left(\frac{x}{2}\right)+w_k}, \label{eq:xrd_sigma}\\
  \Omega_k(x)
  &= A_{\mathrm{asym}}(x;\alpha_k,\mu_k)
  \left\{s_k\frac{1}{\cos\!\left(\frac{x}{2}\right)}+t_k\tan\!\left(\frac{x}{2}\right)\right\}, \label{eq:xrd_omega}
\end{align}
where the trigonometric arguments are converted to radians in the implementation.
The asymmetry factor is
\begin{equation}
  A_{\mathrm{asym}}(x;\alpha_k,\mu_k)=
  \begin{cases}
    \alpha_k & (x\ge \mu_k),\\
    1 & (x<\mu_k),
  \end{cases}
  \qquad (\alpha_k>0).
  \label{eq:xrd_asym}
\end{equation}
The background is modeled as
\begin{equation}
  B(x;\Theta_B)=a\,V\!\left(x;0,\sigma_{\mathrm{bg}},\sigma_{\mathrm{bg}},r_{\mathrm{bg}}\right)+b,
  \label{eq:xrd_bg}
\end{equation}
with $\Theta_B=\{a,\sigma_{\mathrm{bg}},r_{\mathrm{bg}},b\}$.
The full parameter set is $\Theta = \{A_k, \Delta 2\theta_k, r_k, \alpha_k, u_k, v_k, w_k, s_k, t_k\}_{k=1}^{K}\cup\Theta_B$.
Peak positions are determined from crystallographic reference positions $\mu_{k,\mathrm{ref}}$ via a common shift parameter: $\mu_k = \mu_{k,\mathrm{ref}}+\Delta 2\theta_k$.

For artificial data, the noise model is Poisson; the corresponding negative log-likelihood (up to an additive constant) is
\begin{equation}
  E_{\mathrm{Poisson}}(\Theta) = \frac{1}{N}\sum_{i=1}^{N}\left[f(x_i, \Theta) - y_i \ln f(x_i, \Theta)\right].
  \label{eq:xrd_poisson}
\end{equation}
For real data, we use a Gaussian approximation to Poisson counting statistics (mean equal to variance):
\begin{equation}
  p(y\mid x, \Theta) = \frac{1}{\sqrt{2\pi f(x, \Theta)}}\exp\!\left(-\frac{(y-f(x, \Theta))^2}{2f(x, \Theta)}\right),
\end{equation}
with the corresponding negative log-likelihood
\begin{equation}
  E_{\mathrm{Gauss}}(\Theta) = \frac{1}{N}\sum_{i=1}^{N}\left[\frac{1}{2}\ln\!\left(2\pi f(x_i, \Theta)\right) + \frac{(y_i-f(x_i, \Theta))^2}{f(x_i, \Theta)}\right].
\end{equation}

Table~\ref{tab:xrd_priors} lists the prior distributions.
Gamma priors enforce positivity of $\alpha_k$, $u_k$--$w_k$, $s_k$, $t_k$, $A_k$, $a$, and $\sigma_\mathrm{bg}$.
The functional forms are shared between the artificial-data model ($K{=}3$ phases: rutile, anatase, brookite) and the real-data model ($K{=}2$ phases: rutile, anatase); only the data-dependent hyperparameters ($y_\mathrm{max}$, $y_\mathrm{min}$) change.

\begin{table}
  \caption{\label{tab:xrd_priors}Prior distributions for the XRD model. $\mathrm{Ga}(\alpha,\beta)$: gamma (shape $\alpha$, rate $\beta$); $\mathcal{N}(\mu,\sigma^2)$: normal; $\mathrm{U}(a,b)$: uniform.}
  \begin{ruledtabular}
  \begin{tabular}{ccc}
    Parameter & Description & Prior \\
    \hline
    \multicolumn{3}{l}{Crystalline-phase parameters (each phase $k=1,\ldots,K$)} \\
    $\Delta 2\theta_k$ & Peak position shift & $\mathcal{N}(0, 0.05^2)$ \\
    $\alpha_k$         & Asymmetry factor    & $\mathrm{Ga}(5, 4)$ \\
    $r_k$              & Mixing ratio        & $\mathrm{U}(0, 1)$ \\
    $u_k$              & Caglioti U          & $\mathrm{Ga}(1, 10)$ \\
    $v_k$              & Caglioti V          & $\mathrm{Ga}(1, 10)$ \\
    $w_k$              & Caglioti W          & $\mathrm{Ga}(2, 20)$ \\
    $s_k$              & Lorentzian width (sec) & $\mathrm{Ga}(2, 20)$ \\
    $t_k$              & Lorentzian width (tan) & $\mathrm{Ga}(1, 10)$ \\
    $A_k$              & Peak intensity      & $\mathrm{Ga}(4, 4/(y_{\max}-y_{\min}))$ \\
    \hline
    \multicolumn{3}{l}{Background parameters} \\
    $a$                     & Amplitude    & $\mathrm{Ga}(2, 1/y_{\max})$ \\
    $\sigma_{\mathrm{bg}}$  & Width        & $\mathrm{Ga}(2, 0.4)$ \\
    $r_{\mathrm{bg}}$       & Mixing ratio & $\mathrm{U}(0, 1)$ \\
    $b$                     & Offset       & $\mathrm{U}(y_{\min}{-}\sqrt{y_{\min}},\; y_{\min}{+}\sqrt{y_{\min}})$ \\
  \end{tabular}
  \end{ruledtabular}
\end{table}

The rutile sample was titanium(IV) oxide nanopowder (particle size $<$100\,nm, 99.5\% trace metals basis, Sigma-Aldrich, Product No.\ 637262).
The anatase sample was titanium(IV) oxide nanopowder (particle size $<$25\,nm, 99.7\% trace metals basis, Sigma-Aldrich, Product No.\ 637254, CAS 1317-70-0).
XRD measurements were performed using a Rigaku SmartLab instrument (Cu tube, HyPix-3000 multidimensional detector).

\section{\label{sec:appendix_xps}XPS data and model formulation}

X-ray photoelectron spectroscopy (XPS) determines the chemical states of elements by analyzing the kinetic energies of photoelectrons emitted upon X-ray irradiation.
The measured data consist of photoelectron intensities as a function of binding energy.

We model the spectrum with pseudo-Voigt peak functions and a dynamic Shirley background~\cite{Shirley1972_AuValenceBands, HerreraGomez2014_PracticalBackgrounds}.
The forward model is
\begin{equation}
\begin{aligned}
  f(x, \Theta)
  &= \sum_{k=1}^{K} A_k \Biggl[
    \eta_k \exp\!\left(
      -\frac{(x-\mu_k)^2}{2\left(\frac{\sigma_k}{\sqrt{2\ln 2}}\right)^2}
    \right) \\
  &\quad  + (1-\eta_k)\frac{\sigma_k^2}{\sigma_k^2 + (x-\mu_k)^2}
  \Biggr]
  + \operatorname{Shirley}(x, a, b),
\end{aligned}
\label{eq:xps_forward_model}
\end{equation}
where $a$ and $b$ are parameters representing the baseline endpoint heights.
The model parameters are
\begin{equation}
  \Theta = \{A_k, \eta_k, \mu_k, \sigma_k\}_{k=1}^{K} \cup \{a, b\}.
\end{equation}
Following Nagai \textit{et al.}~\cite{Nagai2020_ExchangeMC}, the noise variance depends on the model prediction $f_i=f(x_i,\Theta)$:
\begin{equation}
  \sigma_i^2 = \sigma_0^2 f_i + \sigma_1^2 f_i^2 + \sigma_2^2,
\end{equation}
where $\sigma_0$, $\sigma_1$, and $\sigma_2$ are fixed noise parameters set in this work to $\sigma_0{=}1.0$, $\sigma_1{=}0.01$, and $\sigma_2{=}0.0$.
The $\sigma_0^2 f_i$ term accounts for Poisson counting statistics, and the $\sigma_1^2 f_i^2$ term represents intensity-proportional systematic uncertainty.
The corresponding negative log-likelihood is
\begin{equation}
  E(\Theta) = \frac{1}{N}\sum_{i=1}^{N}\left[\frac{1}{2}\ln(2\pi\sigma_i^2) + \frac{(y_i - f_i)^2}{\sigma_i^2}\right],
\end{equation}
where $y_i$ is the measured photoelectron intensity.

Table~\ref{tab:xps_priors} lists the prior distributions; all are uniform, with the ranges for $A_k$, $a$, and $b$ depending on the observed data.

\begin{table}
  \caption{\label{tab:xps_priors}Prior distributions for the XPS model. $\mathrm{U}(a,b)$: uniform distribution. $y_{\min}$, $y_{\max}$: minimum and maximum observed intensities; $x_{\min}$, $x_{\max}$: binding-energy measurement range; $y_{\mathrm{first}}$, $y_{\mathrm{last}}$: intensities at the start and end of the spectrum.}
  \begin{ruledtabular}
  \begin{tabular}{ccc}
    Parameter & Description & Prior \\
    \hline
    \multicolumn{3}{l}{Peak parameters (each peak $k=1,\ldots,K$)} \\
    $A_k$      & Amplitude    & $\mathrm{U}(\max(0, 0.3\,y_{\min}), 1.05\,y_{\max})$ \\
    $\mu_k$    & Center       & $\mathrm{U}(x_{\min}, x_{\max})$ \\
    $\sigma_k$ & HWHM         & $\mathrm{U}(0.1, 15.0)$ \\
    $\eta_k$   & Mixing ratio & $\mathrm{U}(0, 1)$ \\
    \hline
    \multicolumn{3}{l}{Shirley background parameters} \\
    $a$ & Start intensity & $\mathrm{U}(0.95\,y_{\mathrm{first}}, 1.01\,y_{\mathrm{first}})$ \\
    $b$ & End intensity   & $\mathrm{U}(0.95\,y_{\mathrm{last}}, 1.01\,y_{\mathrm{last}})$ \\
  \end{tabular}
  \end{ruledtabular}
\end{table}

Table~\ref{tab:xps_basis_assignment} assigns each basis function in the $K{=}7$ model to features in the Ni~2p HAXPES spectrum of Ni$_3$Al$_2$O$_3$~\cite{Longo_DeepCoreLevelHAXPES_2023} (840 data points, binding energy $\approx$845--887\,eV).
Assignments for metallic Ni and NiO-like Ni$^{2+}$ components are based on the Ni~2p spin-orbit splitting ($\approx$17.3\,eV) being reproduced by the estimated peak positions.
Peaks $k{=}3$, $4$, and $7$ are attributed to satellites arising from final-state effects (shake-up, plasmon loss) and do not represent independent chemical states.
Reference positions are taken from Refs.~\cite{ThermoFisher_Ni_XPS, Grosvenor2006_NiXPS, Nesbitt2000_Ni2pXPS}.

\begin{table}
  \caption{\label{tab:xps_basis_assignment}Assignment of each basis function to the Ni~2p spectrum for the real XPS data ($K{=}7$ model). Estimated positions and 95\% credible intervals were computed from the REMC(CPU) posterior samples (averaged over 100 runs).}
  \begin{ruledtabular}
  \begin{tabular}{clccc}
    $k$ & Assignment & Category & Est.\ pos.\ [95\%\,CI] (eV) & Ref.\ (eV) \\
    \hline
    1 & Ni$^{0}$ 2p$_{3/2}$       & Main      & 851.65\,[851.61, 851.69] & $\sim$852.6 \\
    2 & Ni$^{2+}$ 2p$_{3/2}$      & Main      & 853.44\,[853.28, 853.61] & 853.7--854.6 \\
    3 & 2p$_{3/2}$ satellite      & Satellite & 858.83\,[858.56, 858.99] & $\sim$858 \\
    4 & 2p$_{3/2}$ high-BE        & Satellite & 863.04\,[859.34, 864.07] & $\sim$861 \\
    5 & Ni$^{0}$ 2p$_{1/2}$       & Main      & 869.07\,[868.93, 869.22] & $\sim$869.9 \\
    6 & Ni$^{2+}$ 2p$_{1/2}$      & Main      & 870.86\,[870.63, 871.14] & $\sim$871.0 \\
    7 & 2p$_{1/2}$ satellite      & Satellite & 877.02\,[876.87, 877.18] & $\sim$878 \\
  \end{tabular}
  \end{ruledtabular}
\end{table}

\end{document}